\documentclass[prb,twocolumn,superscriptaddress,amsmath,amssymb,showpacs,floatfix,preprintnumbers]{revtex4-1}
\usepackage{amsmath}
\usepackage{amssymb}
\usepackage{graphicx}
\usepackage[export]{adjustbox}
\usepackage{epstopdf}
\usepackage{color}
\usepackage{epsfig}
\usepackage{amsmath}
\usepackage[colorlinks=true,citecolor=blue,linkcolor=blue]{hyperref}
\usepackage{amssymb}

\DeclareMathAlphabet{\bi}{OML}{cmm}{b}{it}

\begin{document}

\title{An electrical dipole on gapped graphene: bound states and atomic collapse}
\author{R. Van Pottelberge}\email{robbe.vanpottelberge@uantwerpen.be}
\affiliation{Departement Fysica, Universiteit Antwerpen, Groenenborgerlaan 171, B-2020 Antwerpen, Belgium}

\author{B. Van Duppen}\email{ben.vanduppen@uantwerpen.be}
\affiliation{Departement Fysica, Universiteit Antwerpen, Groenenborgerlaan 171, B-2020 Antwerpen, Belgium}

\author{F. M. Peeters}\email{francois.peeters@uantwerpen.be}
\affiliation{Departement Fysica, Universiteit Antwerpen, Groenenborgerlaan 171, B-2020 Antwerpen, Belgium}
\begin{abstract}
We investigate the energy spectrum, wave functions and local density of states (LDOS), of an electrical dipole placed on a sheet of gapped graphene as function of the charge strength $Z\alpha$, for different sizes of the dipole and for different regularisation parameters. The dipole is modeled as consisting of a positive and negative charge. Bound states are found within the gap region with some energy levels that anti-cross and others that cross as function of the impurity strength $Z\alpha$. The anti-crossings are more pronounced and move to higher charges $Z\alpha$ when the length of the dipole decreases. These energy levels turn into atomic collapse states when they enter the positive (or negative) energy continuum. A smooth transition from the single impurity behavior to the dipole one is observed: the states diving towards the continuum in the single impurity case are gradually replaced by a series of anticrossings that represent a continuation of the diving states in the single impurity case. By studying the local density of states at the edge of the dipole we show how the series of anticrossings persist in the positive and negative continuum.
\end{abstract}

\maketitle

\section{Introduction}
One of the most intensively studied stable two-dimensional (2D) systems is monolayer graphene (MLG). This is not only due to its very interesting thermal and electrical properties of MLG, but also because of its ability to mimic relativistic physics at low energies. Interesting phenomena such as Klein tunneling that were predicted almost a century ago [\onlinecite{Klein}], and that escaped experimental verification, have been recently observed in graphene [\onlinecite{Stander}]. 

Placing MLG on a substrate like $\text{Si}\text{C}$ or hexagonal boron nitride (hBN) breaks the sublattice symmetry creating a gapped spectrum. For example, putting MLG on a $\text{Si}\text{C}$ substrate creates a gap of 260 meV [\onlinecite{zhou}]. Adding a Coulomb impurity to the gapped MLG system creates a close analog of a relativistic hydrogen atom. In relativistic physics it was predicted that for extremely high charges of the nucleus ($Z>170$) bound states in the mass gap between the positive and negative continuum approach the positron continuum as function of the nuclear charge and can even hybridize with it as the nuclear charge increases when the bound states enter the continuum. As a consequence turning the bound state into a resonant state with a finite lifetime [\onlinecite{Greiner}]. The diving in the continuum is discussed in Refs. [\onlinecite{Greiner}] and [\onlinecite{Pomeranchuk}] and is referred to as \textit{atomic collapse}, sometimes also the name \textit{supercritical instability} is used. The latter refers to the appearence of an unstable quasi-bound state [\onlinecite{Shytov1}]. 

It is important to mention that the appearance of such a resonant state is a clear signature of atomic collapse. The production of an electron-hole pair as the signature for the atomic collapse depends on the Fermi energy. If the systems Fermi energy is such that the bound state is empty an electron-hole pair will be created upon entering the negative continuum. However, in the case of a filled bound state such a pair creation will not occur [\onlinecite{Pomeranchuk}]

Due to the extremely high charge needed to realise atomic collapse in real 3D atoms it was never observed [\onlinecite{Cowan}]. The discovery of graphene however opened a new window on the research of atomic collapse both in gapless [\onlinecite{Shytov}] and gapped graphene [\onlinecite{pereira}-\onlinecite{Rodin}]. This can be traced back to the fact that the effective fine structure constant is much larger in graphene than in relativistic physics which lowers the charge needed for the observation of atomic collapse from $Z_c\approx 170$ to $Z_c\approx 1$. Recently, atomic collapse resonances were observed for the first time in bulk graphene for: (i) charged calcium dimers placed on a graphene sheet [\onlinecite{crommie}] and (ii) charged vacancies in graphene [\onlinecite{andrei0}]. This raises the question: how will atomic collapse manifest itself in the presence of different arrangements of charges?

\begin{figure}[t]
\includegraphics[scale=0.3]{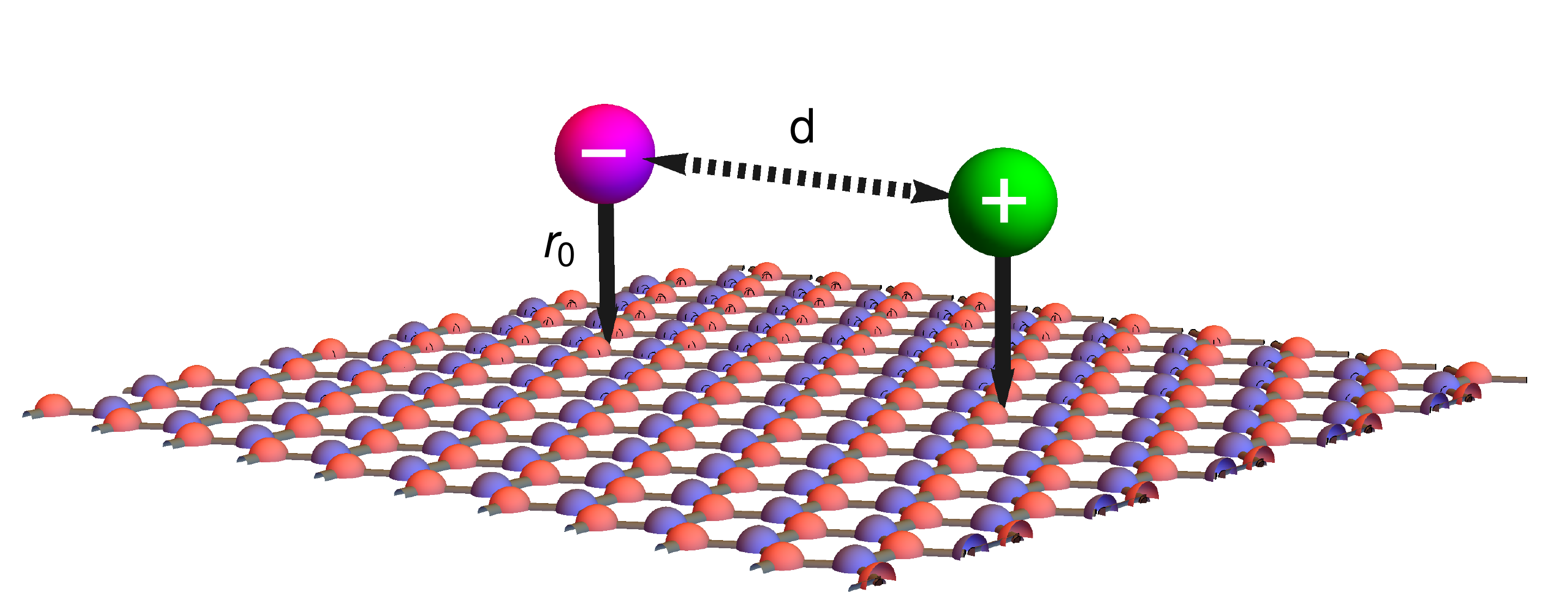}
\centering
\caption{(Color online) A dipole, modeled by two opposite charges, is placed on top of a graphene sheet, is modeled by two opposite charges. The charges are placed at a distance $r_0$ from the graphene sheet (indicated by the black arrows) and are separated from each other by a distance $d$ (indicated by the dashed arrow).}
\end{figure}

A very important case is a dipole configuration which we model as a negatively and positively charged impurity (see Fig. 1). Relativistic particles in the presence of a dipole field are rarely studied due to the absence of very large negative/positive charges [\onlinecite{Martino}]. The discovery of graphene however opens up the possibility to study dipoles in a relativistic context without the need for highly charged impurities. One important question for example that we want to answer is: what will happen to the atomic collapse effect in the case of a dipole? Previous research on this topic was limited to approximate solutions. For example, in Refs. [\onlinecite{Martino}] and [\onlinecite{Martino2}] the problem of Fig. 1 in gapped graphene was studied within the dipole approximation, i.e. the interaction potential was taken as $V(r)\sim \cos{\theta}/r^2$. In Refs. [\onlinecite{Gorbar}] and [\onlinecite{Gorbar2}] the problem was studied using variational and linear combination of orbitals (LCAO) approaches and the investigation was limited to the lowest energy electron and highest energy hole state. Interestingly, regarding the atomic collapse different conclusion can be found in the existing literature. While Ref. [\onlinecite{Martino}] seems to argue that no atomic collapse occurs, Ref. [\onlinecite{Gorbar}] suggests that a new kind of atomic collapse is present in a dipole field. This apparent discrepancy, is our major motivation to present an exact numerical solution of the problem. 

It is important to stress that in this paper by atomic collapse we mean the effect were a bound state enters the negative continuum producing a resonance (i.e. quasi-bound state). Throughout the paper we will also refer to this effect as ``diving into the continuum". The reason to stress this is that in the existing literature sometimes confusing definitions for atomic collapse are used (e.g. see comment [\onlinecite{Zarenia}] and the corresponding reply [\onlinecite{Li}]). 

The purpose of this study is two-fold. First, we investigate the full spectrum of a physical dipole in gapped graphene within the continuum approximation. Note that all previous studies were also done within the continuum approximation. Obtaining the full spectrum will give us more insight in the physics. Also, the spectrum can be accurately determined for all energy levels and arbitrary dipole sizes. We would like to emphasize that such a detailed study does not yet exist in the literature. Determining the exact full spectrum is very useful regarding the experimental observation of atomic collapse. The second purpose of this paper is to investigate how the atomic collapse phenomenon evolves from the single impurity case to the dipole case.

The paper is structured as follows: in Sec. II we present the model and derive the equations that have to be solved numerically. In Sec. III we present our numerical results on bound states inside the gap. The numerical results on atomic collapse states inside the continuum are given in Sec IV, and in Sec V we present our conclusions.  

\section{Model}
In this section we present the mathematical model and equations used throughout this paper.
\subsection{Hamiltonian}
We solve the dipole problem within the continuum model using the full Dirac Hamiltonian:
\begin{equation}
\hat{H}=-i\hbar v_F(\sigma_x\partial_x+\sigma_y\partial_y)+\Delta\sigma_z+V(x,y)I_2.
\end{equation}
Here $\sigma_i$ are the Pauli matrices, $\Delta$ is the size of the gap, $I_2$ is the identity matrix and $V(x,y)$ is an external electrostatic potential. Choosing the $x$ axis along the positive and negative charges (see Fig. 1) we obtain the following form for the physical dipole potential:
\begin{equation}
V(x,y)=\frac{Z\alpha_0}{\sqrt{(x+\frac{d}{2})^2+y^2+r_0^2}}-\frac{Z\alpha_0}{\sqrt{(x-\frac{d}{2})^2+y^2+r_0^2}}.
\end{equation}
Here $\alpha_0=e^2/4\pi\kappa$, $Z$ is the charge number, $\kappa$ the effective dielectric constant, $d$ the distance between the two charges and $r_0$ the distance from the charges to the graphene sheet. The experimental situation is shown schematically in Fig. 1. By taking into account the distance from the charges to the graphene sheet represents a more physical situation as compared to point charges situated inside graphene, it is also necessary to solve the problem for all values of the charge of the two impurities [\onlinecite{Zarenia}]. The effect of this regularization (i.e. taking $r_0\neq 0$) is discussed more extensively in appendix A.

Following Ref. [\onlinecite{Martino}] we will work in units of $\hbar v_F$ ($\approx 658 \text{ meVnm}$) giving the following set of differential equations we need to solve: 
\begin{subequations}
\begin{equation}
\begin{split}
\left(-i\frac{\partial}{\partial x}-\frac{\partial}{\partial y}\right)\psi_b+\bar{\Delta}\psi_a+\bar{V}(x,y)\psi_a=\bar{E}\psi_a,
\end{split}
\end{equation}
\begin{equation}
\left(-i\frac{\partial}{\partial x}+\frac{\partial}{\partial y}\right)\psi_a-\bar{\Delta}\psi_b+\bar{V}(x,y)\psi_b=\bar{E}\psi_b.
\end{equation}
\end{subequations}
Here we introduced the rescaled energy $\bar{E}=E/\hbar v_F$ and band gap $\bar{\Delta}=\Delta/\hbar v_F$. $V(x,y)$ is given by the expression:
\begin{equation}
\bar{V}(x,y)=\frac{Z\alpha}{\sqrt{\left(x+\frac{d}{2}\right)^2+y^2+\bar{r}_0^2}}-\frac{Z\alpha}{\sqrt{\left(x-\frac{d}{2}\right)^2+y^2+\bar{r}_0^2}}.
\end{equation}
Where we introduced the dimensionless effective fine structure constant $\alpha=\alpha_0/\hbar v_F$.

Note that the band gap itself leads to a characteristic length scale for the states located inside the gap given by the effective Compton wavelength $R_\Delta=\hbar v_F/\Delta$. 

\subsection{Total angular momentum}
Before examining numerical solutions we would like to emphasize that the two-impurity problem cannot be solved exactly. There is no longer cylindrical symmetry and as a consequence the total angular quantum number $\hat{J}_z=\hat{L}_z+\frac{\hbar}{2}\sigma_z$ is no longer a conserved quantity. This is in contrast with the single negative/positive impurity case where $\hat{J}_z$ is conserved and the expectation value takes the discrete values $<\hat{J}_z>/\hbar=(m+1/2)$, where $m$ can take the values $0,\pm 1,\pm 2,...$. However, it is still interesting to calculate the expectation value of $\hat{J}_z$ which in Cartesian Coordinates is given by:
\begin{equation}
\hat{J}_z=-i\hbar \left(x\frac{\partial}{\partial y}-y\frac{\partial}{\partial x}\right)I_2+\frac{\hbar}{2}\sigma_z.
\end{equation}
The expectation value of the total angular momentum for a state described by the two component wave function $\Psi=(\psi_a,\psi_b)$ is then given by the following expression:
\begin{equation}
\begin{split}
&\frac{<\bar{L}_z>}{\hbar}=-i\int\psi_a^*\left(x\frac{\psi_a}{\partial y}-y\frac{\psi_a}{\partial x}\right)dxdy \\ &-i\int\left[\psi_b^*\left(x\frac{\partial\psi_b}{\partial y}-y\frac{\partial\psi_b}{\partial x}\right)+\int\left(\mid\psi_a \mid^2-\mid\psi_b \mid^2\right)\right]dxdy
\end{split}
\end{equation}

\subsection{Finite size simulation}
As discussed in the introduction, atomic collapse occurs when bound states dive into the continuum turning them into a resonance. However in light of atomic collapse we need information about the continuum that will show us unambiguously that the diving series of anticrossings persists into the continuum. In order to get an idea about these extended states inside the continuum we consider a finite size system which we solve using the finite elements method. Such an approach was already suggested by Pomeranchuck [\onlinecite{Pomeranchuk}]. The finite size is simulated by introducing an infinite mass potential at a distance $R$ (in nm) by replacing $\bar{\Delta}$ in Eq. (1) by the following expression: 
\begin{equation}
\begin{split}
\bar{\Delta}\rightarrow \bar{\Delta}+\bar{\Delta}_0\Big[\tanh{\left(\frac{x-R}{c}\right)}+\tanh{\left(\frac{y-R}{c}\right)}+\\ \tanh{\left(-\frac{x+R}{c}\right)}+\tanh{\left(-\frac{y+R}{c}\right)}+4\Big].
\end{split} 
\end{equation}
Here, $\Delta_0$ is the height of the infinite mass potential outside the confinement region, $R$ is the size of the flake and $c$ defines the sharpness of the edge. Note that imposing an infinite mass boundary condition does not take into account any edge specifics and in principle the edge of the flake consists of a mixture of zig-zag and armchair boundaries [\onlinecite{Beenakker}]. However, in this work edge effects are not of relevance and we will therefore ignore them by taking $R$ sufficiently large such that the finite size does not influence our conclusions.   

\begin{figure*}[t]
\includegraphics[trim={3.8cm 0cm 3.4cm 0},clip,scale=0.405]{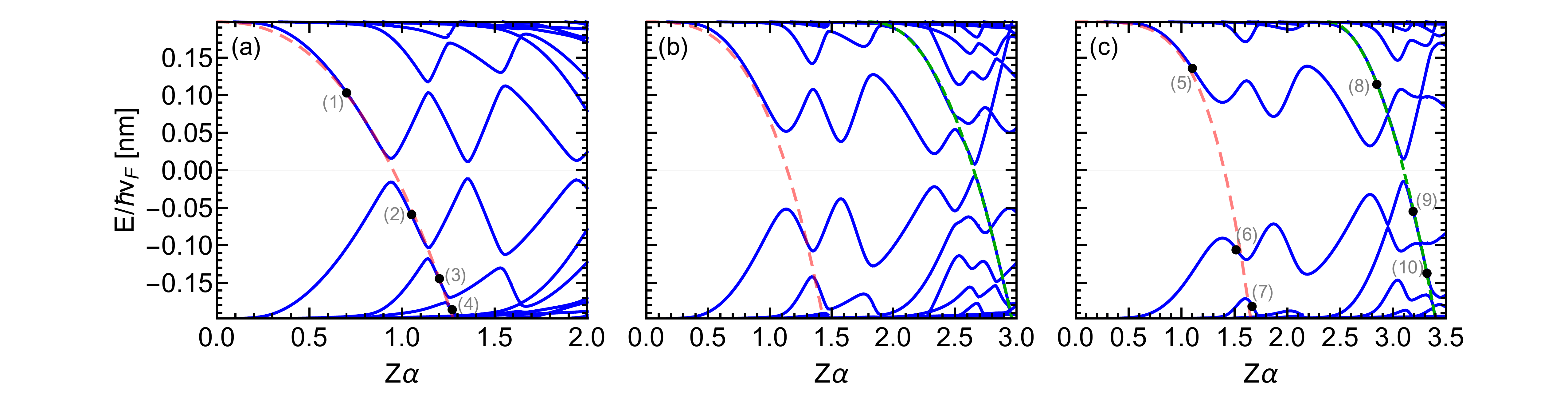}
\centering
\caption{(Color online) The dipole spectrum as function of the charge $Z\alpha$ of the impurities for a gap $\Delta=130$ meV and for three distances between the impurities: (a) 10 nm, (b) 5 nm and (c) 3 nm. In all figures we used the same regularisation parameter $r_0=0.4$ nm. The red curves are fits with the expression $\bar{E}=a(Z\alpha)^b+\Delta/\hbar v_F$. The green curves are fits with the expression $\bar{E}=\Delta/\hbar v_F+a(Z\alpha-b)^c$.}\label{fig:fig1}
\end{figure*}

\begin{figure}
\includegraphics[scale=0.32]{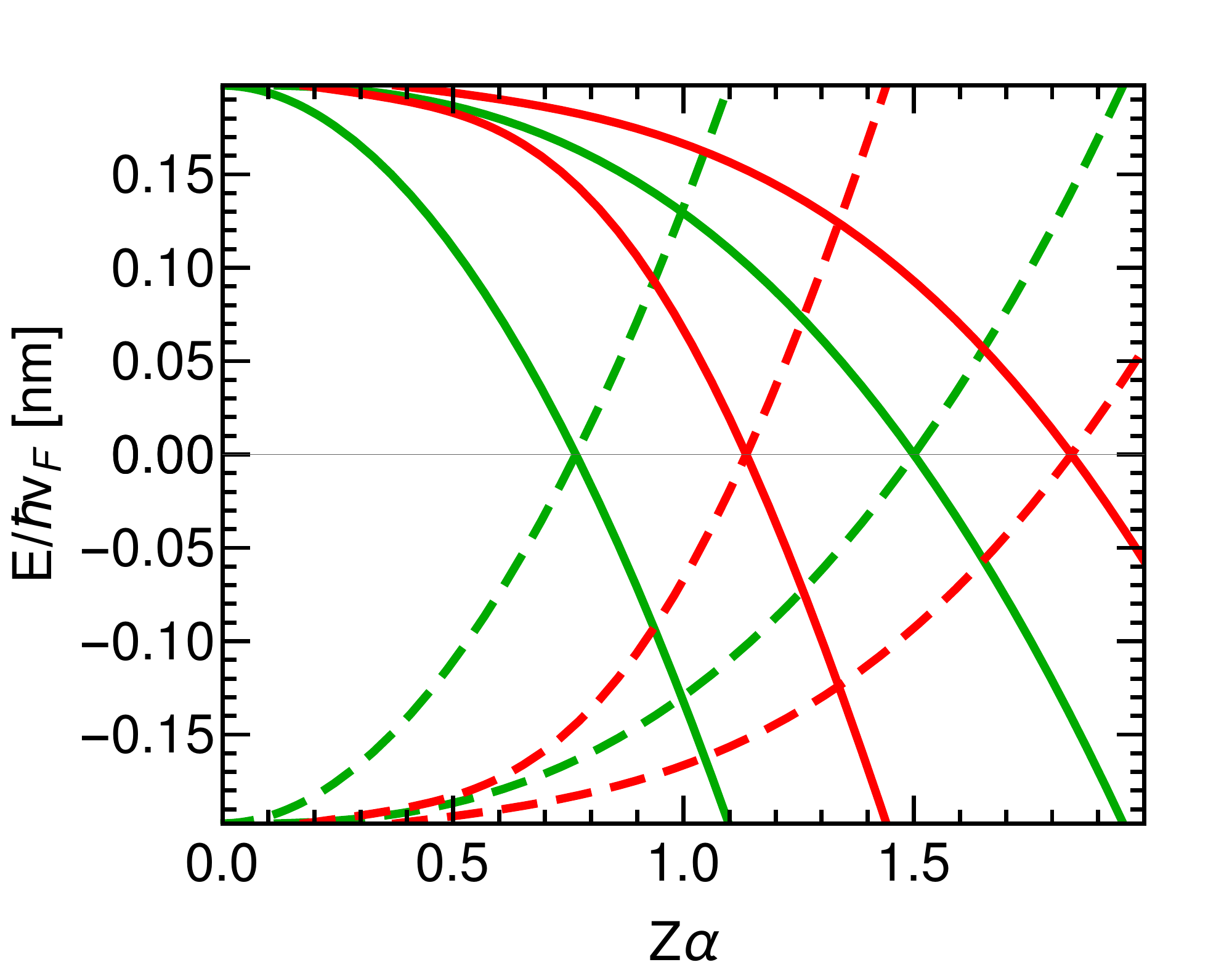}
\caption{(Color online) Single impurity spectrum for positive (solid curves) and negative charges (dashed curve) as function of the charge $Z\alpha$. The states with total angular momentum 0.5 ($m=0$ states) are shown in green and the states with total angular momentum -0.5 ($m=-1$ states) are shown in red.}
\end{figure}

\section{Bound states in the gap}
In this subsection we focus on the bound state spectrum located inside the gap. We will show that the diving of states towards the continuum seen in the single impurity case [\onlinecite{Pereira}] persists in the case of a dipole as a series of diving anticrossings. We will show that with decreasing inter-charge distance the anticrossings become more profound but their diving into the continuum persists. 
\subsection{Changing inter-charge distance}
We solve the set of coupled equations (3a) and (3b) using the finite elements method implemented in Mathematica, which discretizes the lattice and allows for a numerical solution of the problem. It should be mentioned that the discretization procedure introduces spurious solutions due to the so called \textit{fermion doubling problem} [\onlinecite{Chandra}]. However, by studying the dependence of the energy levels on the mesh size and the behavior of the wave functions we are able to identify and eliminate these spurious solutions. We will start by considering the dependence of the full spectrum on the impurity strength for different distances between the impurities. 

In Fig. \ref{fig:fig1} the energy spectrum of a dipole in gapped graphene is shown as function of the impurity strength $Z\alpha$ for three different dipole sizes (10, 5 and 3 nm). The size of the gap is taken $\Delta=130$ meV which corresponds to the gap created by putting graphene on a $SiC$ substrate [\onlinecite{zhou}]. This gap results in a Compton wavelength $R_\Delta=5$ nm. In all calculations we used the regularisation $r_0=0.4$ nm which is the distance between the graphene sheet and the charges and is an experimentally realistic value, see Refs. [\onlinecite{andrei}, \onlinecite{crommie}]. A small discussion on the effect of the regularization parameter on the spectrum is provided in Appendix A. 

Let's first discuss Fig. \ref{fig:fig1}(a) where we show the spectrum for two charges seperated by 10 nm from each other, which is large compared to the Compton wavelength. When the charge of the impurities increases electron states start to descend from the positive continuum due to the positively charged impurity while hole states start to rise from the negative continuum due to the negatively charged impurity. Note that the first bound state is already allowed to sink into the gap for arbitrary small values of the charges in accordance with what was shown in Ref. [\onlinecite{Martino}]. It should be mentioned that this is in sharp contrast with 3D non-relativistic physics where bound states are only found after a certain critical value of the dipole moment [\onlinecite{connolly}]. When the charge is further increased the electron states sink further in the gap while the hole levels do the exact opposite, until around $Z\alpha\approx 0.9$ where the levels anticross with each other. The approaching electron and hole levels exhibit the same symmetry, which prohibits the levels from crossing [\onlinecite{Wigner}]. If we ignore the anti-crossing for one moment we notice that both levels will continue and for example the electron states dive towards the negative continuum while hole states do the exact opposite.

In Fig. 3 we plotted the $m=0$ (green curves) and $m=-1$ (red curves) single impurity states on top of the dipole spectrum shown in Fig. 2(a). From this figure it can be readily seen that the diving of the energy levels as function of the impurity strength $Z\alpha$ is replaced by the diving through a series of anticrossings. Notice that for the single impurity case electron states dive into the continuum at a smaller $Z\alpha$ value as compared to the dipole case. This can be explained from the fact that the tail of the single impurity Coulomb potential is shortened in the dipole case effectively reducing it's strength. 

In Fig. 2 the first electron bound state was fitted to the function $\bar{E}=\Delta/\hbar v_F+a(Z\alpha)^b$ and shown as dashed (red) curves in Fig. 2. We found respectively for $d=10, 5$ and $3$ nm the following values for the parameters $\{a,b\}=\{-0.22,2.36\}$, $\{-0.13,3.07\}$ and $\{-0.05,4.27\}$. If we restrict the fitting to small $Z\alpha$-values we obtained: $\{a,b\}=\{-0.26,2.86\}$, $\{-0.13,3.65\}$ and $\{-0.04,4.47\}$. For the first single impurity state shown in Fig. 3 we find that the functional behavior can be exactly fitted with the coefficients $\{a,b\}=\{-0.33,1.93\}$. It is clear from our results that the coefficients increase with decreasing inter-charge distance. Note that for perfect Rydberg like behavior one would obtain $b=2$. 

\begin{figure}[t]
\includegraphics[scale=0.40]{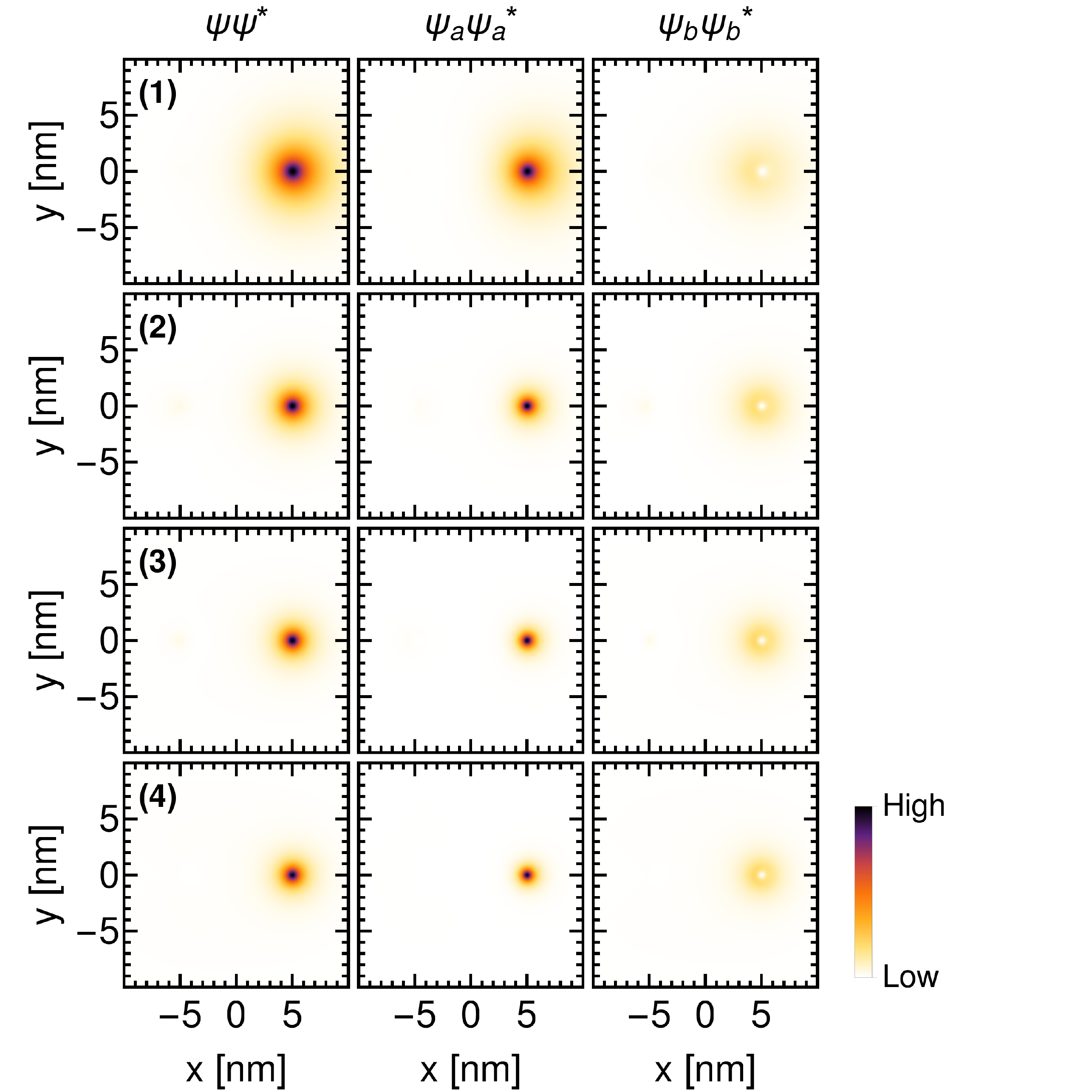}
\centering
\caption{(Color online) Total probability density and densities on each of the sublattices for the points marked in Fig. 2(a).}
\end{figure}

The continuation of the levels regardless of the anticrossings is further supported by Fig. 4 where we plotted the total probability as well as the probability on the two separate sublattices for the points indicated in Fig. 2(a). For $Z\alpha=0.5$ (point (1)) the wave function is clearly localized at the positively charged impurity located at $x=5$ nm. When the charge increases the state becomes more localized closer to the positively charged impurity, as expected. However, it is clear that the behavior of the state remains unchanged after the anticrossing, which can be seen from the densities shown for the points (2)-(4). The state keeps its electron character and localizes closer and closer towards the positively charged impurity with increasing charge. 

Note also from Fig. 4 that most of the probability density is located on the A sublattice, and that only a smaller ring shaped portion of the probability density is located on the B sublattice. This can be explained from the fact that the gap term breaks the sublattice symmetry leading to a small and large component.  

\begin{figure}[t]
\includegraphics[trim={1.5cm 0cm 3.8cm 0},clip,scale=0.44]{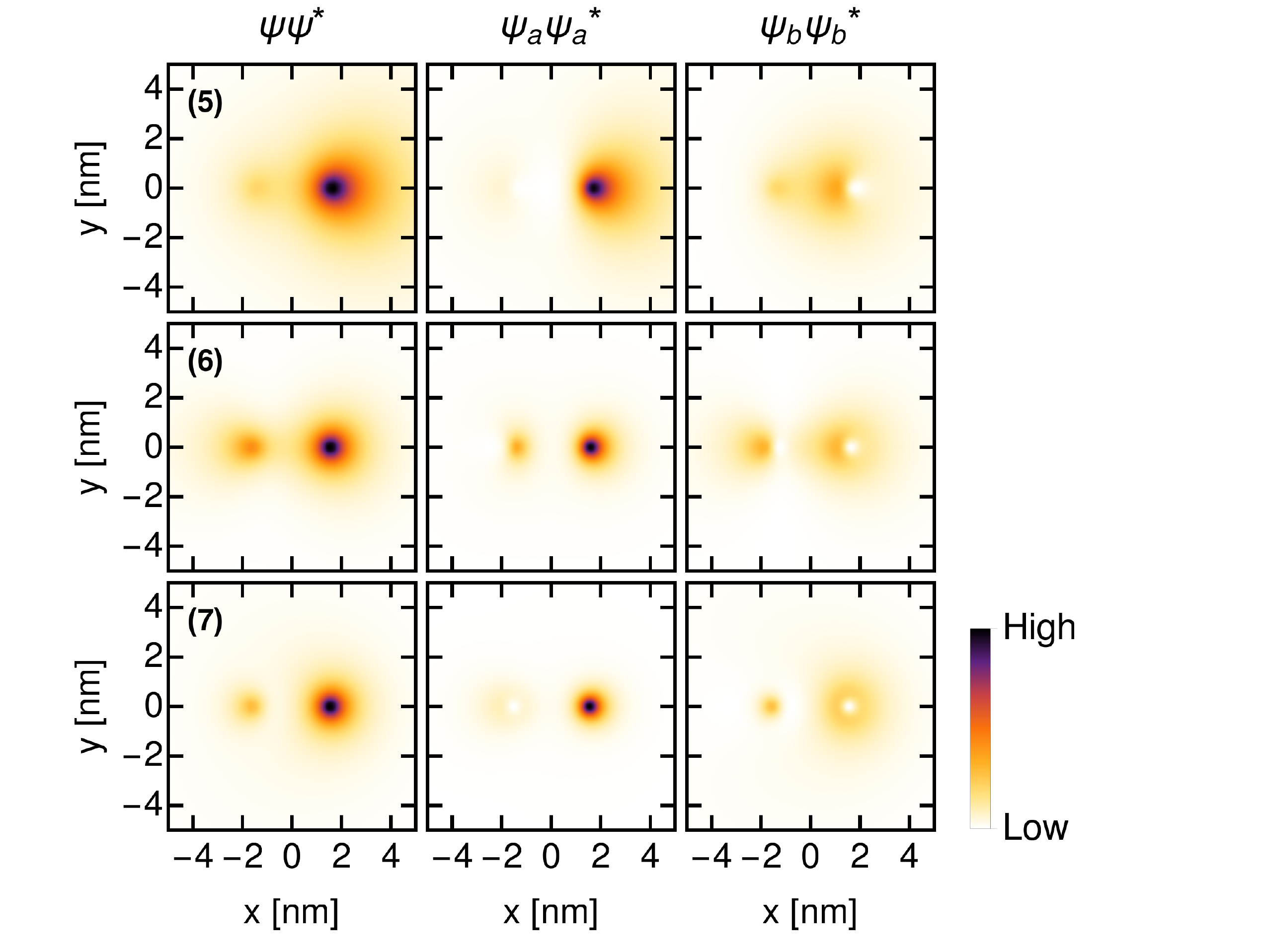}
\centering
\caption{(Color online) Total probability density and densities on each of the sublattices for the points marked in Fig. 2(c).}
\end{figure}

\begin{figure}[t]
\includegraphics[trim={1.5cm 0cm 3.8cm 0},clip,scale=0.44]{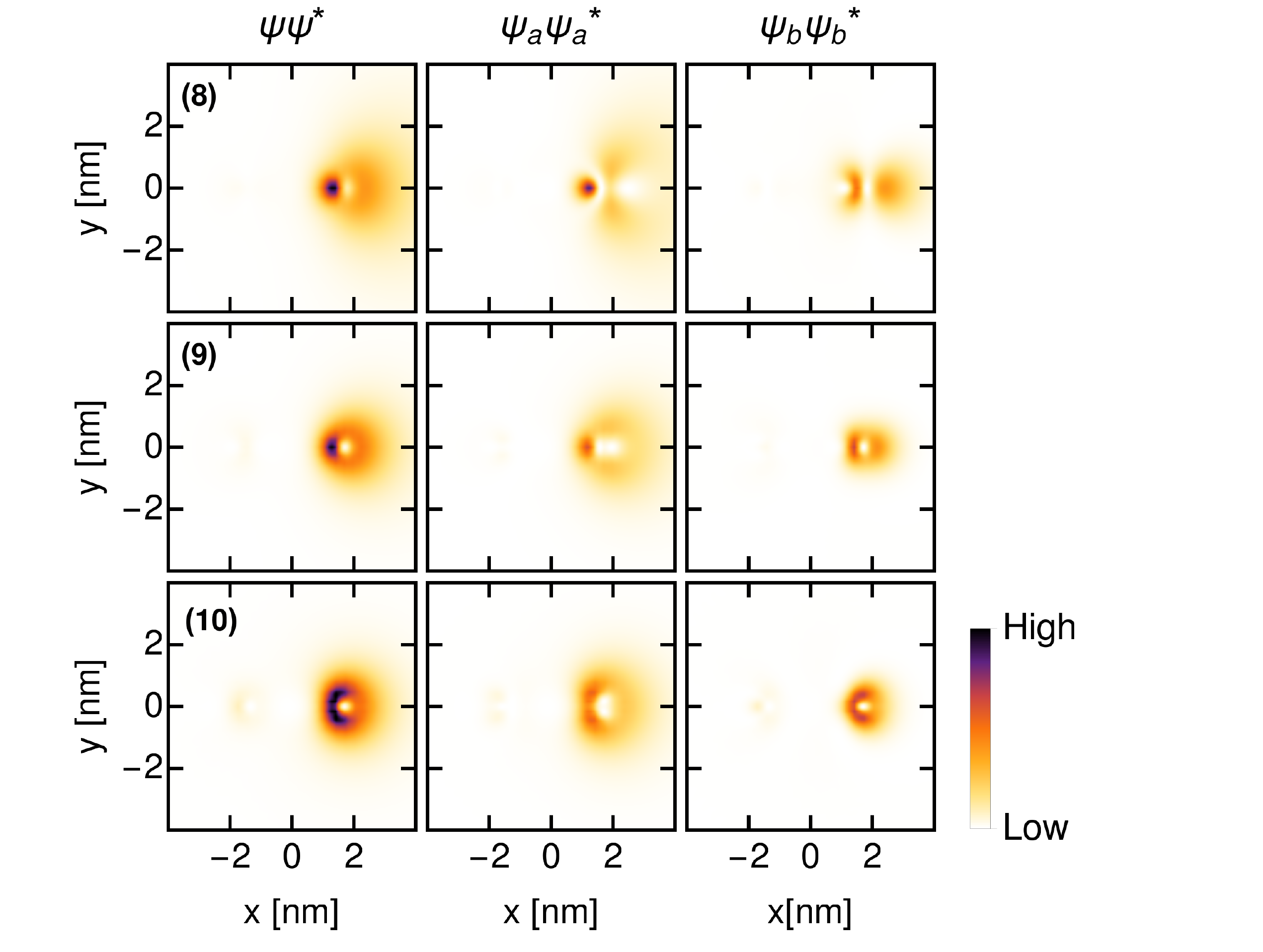}
\centering
\caption{(Color online) Total probability density and densities on each of the sublattices for the points marked in Fig. 2(c).}
\end{figure}

Up to now we focused on the dipole spectrum of two charges located at a fairly large distance from each other, indeed for $d=10$ nm the Compton wavelength $R_\Delta=5$ nm is small compared to the inter-charge distance. Therefore it is interesting to study how the spectrum evolves with decreasing inter-charge distance. In Figs. 2(b) and (c) we show the dipole spectrum for $d=5$ nm and $d=3$ nm, respectively. With decreasing $d$ the anticrossings become more profound. This is due to the fact that the charges are located closer to each other which increases the overlap between the electron and hole wave functions. However, exactly as in Fig. 4 the continuation of the energy levels regardless of the anticrossings can be clearly observed when plotting the probability densities for the points marked in Fig. 2(c) as done in Fig. 5. For the wave functions at point (5) it is clear that the main portion of the density is located at the positive impurity. However, a small portion is also located at the negative impurity, in contrast with Fig. 4, which is due to the fact that the charges are located much closer to each other. When the charge $Z\alpha$ is increased, see points (6) and (7), the probability density clearly localizes more on the positively charged impurity while a small portion is still located at the negatively charged impurity. This behavior shows again that the diving of the state persists regardless of the anticrossings. 

Note that in Fig. 5 most of the probability density is located at the A sublattice, exactly as in Fig. 4. Interestingly any probability density located between the two charges resides on the B sublattice only. The A sublattice has an anti-bonding character while the B sublattices has a bonding character. This can be explained from the fact that the probability density is less localised on the B sublattice compared to the A sublattice (subtly seen in Fig. 4) creating a larger overlap between the A sublattice wave functions. 

Up to now our main focus was on the first series of anticrossings shown in Figs. 2(a)-(c). However from Figs. 2(b) and (c) and partially in (a) another distinct series of anticrossings are observed (around $Z\alpha\approx 3$) that towards the negative/positive continuum and exhibit crossings with some of the levels. In particular in Fig. 2(c) it can be seen that this series of anticrossings behaves in a very distinct way from the other ones. We fitted this energy level by the function $\bar{E}=\Delta/\hbar v_F+a(Z\alpha-b)^c$ which is shown as dotted (green) curves in Fig. 2(b) and (c). We found respectively the values $\{a,b,c\}=\{-0.13,1.80,2.25\}$ for Fig. 2(b) and  $\{a,b,c\}=\{-0.38,2.40,1.97\}$ for Fig. 2(c). 

\begin{figure}
\includegraphics[scale=0.35]{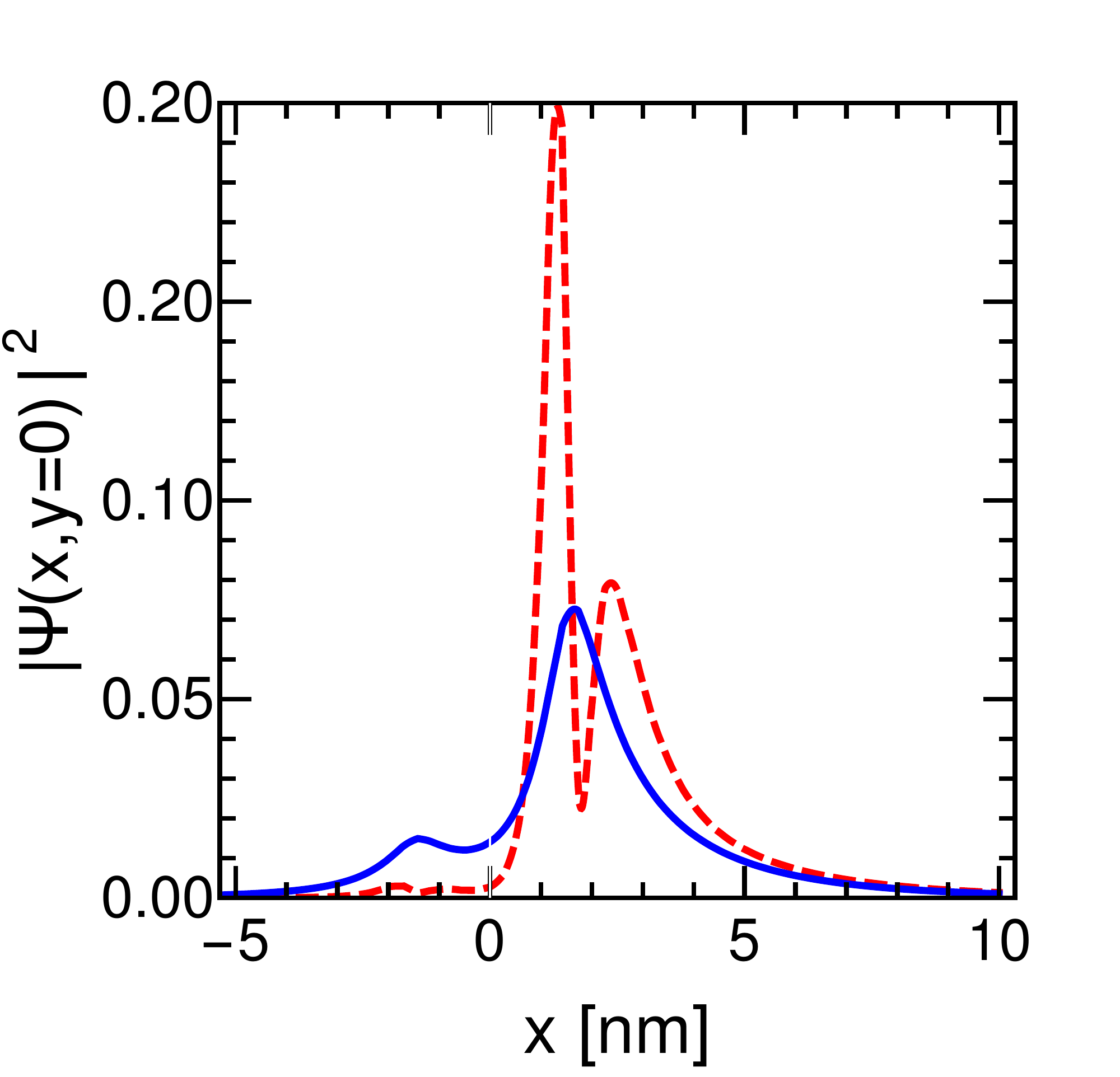}
\caption{Probability density for $y=0$ as function of the $x$ direction for the state (7) shown in Fig. 5 (blue-solid) and the state (10) shown in Fig. 6 (red-dashed).}
\end{figure}

In order to understand this peculiar behavior of this descending state with sharper anticrossings we plot the probability density in Fig. 6 for the points marked in Fig. 2(c). First of all we observe that in contrast with the probability densities shown in Fig. 5 the probability density is almost solely located on the positively charged impurity which explains the sharpness of the anticrossings. Next to that it can be seen that the probability density is highly asymmetric in the $x$ direction compared to those shown in Figs. 4 and 5. The probability density shows a peak next to the positive impurity center in the direction of the negative one and another peak in the probability density is observed right of the positively charged impurity. When the impurity charge increases the probability remains localized on the positive impurity analogous to Figs. 4 and 5 showing again a clear continuation of the energy level regardless of the anticrossings. However, when the charge is increased the general shape of the probability density changes drastically. The probability density gets pushed closer to the positive impurity and starts to aquire a `` donut"-like shape. 

In Fig. 7 we plot the probability in the $x$ direction for the density (5) shown in Fig. 5 and the density (8) shown in Fig. 6. The density of point (5) has a clear peak on the positively charged impurity which is a $1s$-like state behavior. The density of the point (8) on the other side has a dip and another peak next to the impurity on the right side. The dip and extra peak are $2s$-like behavior. The fact that the dip and additional peak are observed on the right side of the impurity is due to the symmetry breaking in the $x$-direction. 

In contrast with the densities shown in Figs. 4 and 5 the densities in Fig. 6 show almost an equal distribution between the A and B sublattices. The probability densities clearly lack angular symmetry around the individual impurities compared to the densities shown in Fig. 4. 

\subsection{Total angular momentum}
In this subsection we calculate the expectation value of the total angular momentum operator which will reinforce our previous interpretation of the anti-crossings. However, the usefulness of the total momentum operator is limited to large inter-impurity distances and smaller charges when the breaking of angular symmetry is less dramatic.

\begin{figure}[h!!!]
\includegraphics[scale=0.43]{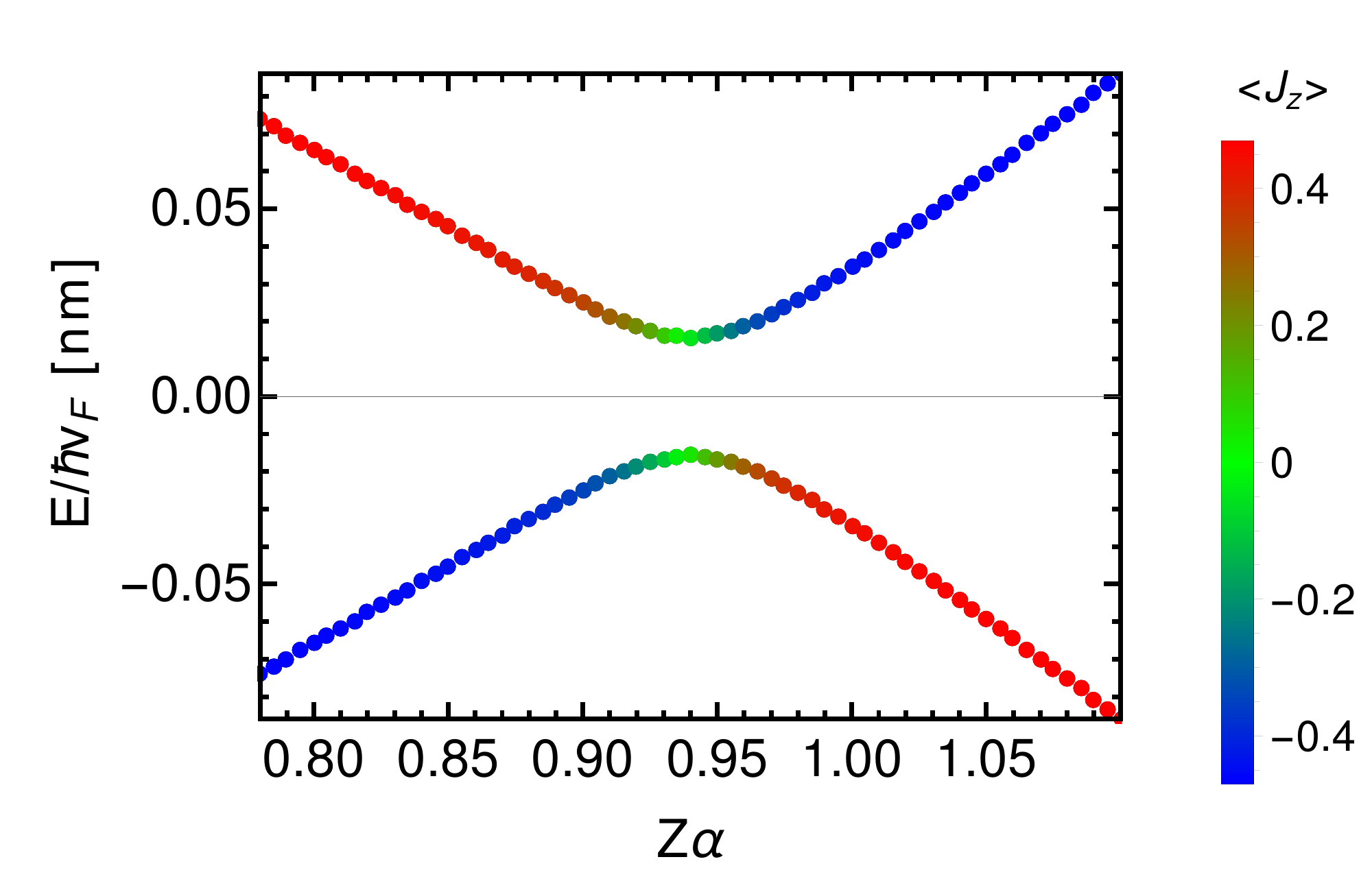}
\includegraphics[trim={0cm 0cm 0cm 0},clip,scale=0.4]{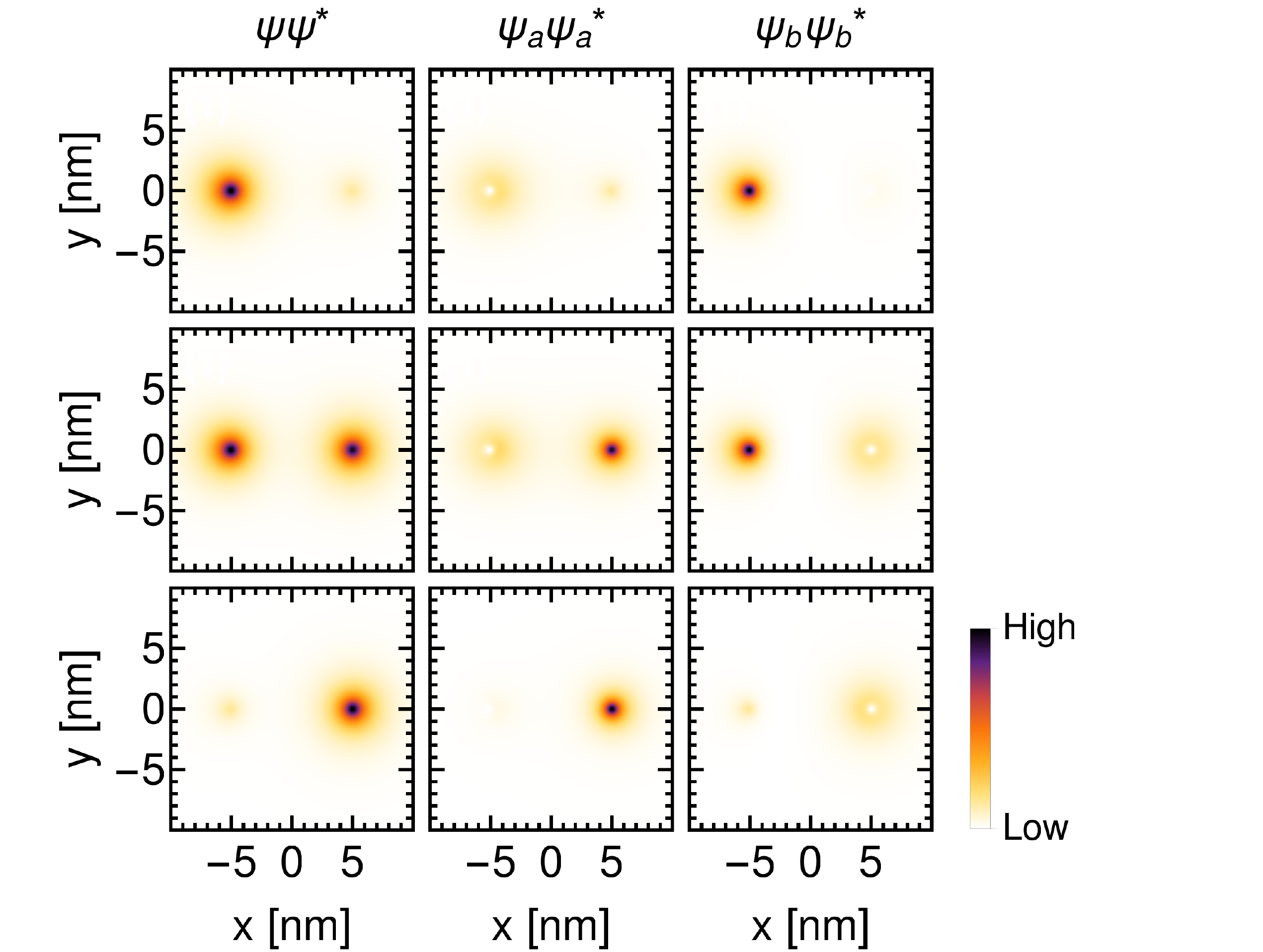}
\caption{Close up of the first anticrossing seen in Fig. 2(a) (between points (1) and (2)). For each point we calculated the expectation value of the angular momentum operator $<\hat{J}_z>$. The bottom figures show the probability density before ($Z\alpha=0.9$, upper panel), during ($Z\alpha=0.94$, middle panel) and after ($Z\alpha=0.98$, bottom panel) the lower anticrossing.}
\end{figure}

It is interesting to calculate the expectation value of the total momentum operator $\hat{J}_z$ for the points shown in Fig. 2(a). For the points (1) to (3) we find for $\{(i),<\hat{J}_z>/\hbar\}$ respectively the values $\{(1),0.48\}$, $\{(2),0.46\}$, $\{(3),0.35\}$. Before the first anticrossing the expectation value of the first diving state is almost 0.5 which corresponds to the $m=0$ level of the single impurity system (i.e. $1s$-state), for the corresponding hole state a value close to the $m=-1$ state is found. This is consistent with the single impurity results shown in Fig. 3 and with the approximate results in Refs. [\onlinecite{Gorbar}] and [\onlinecite{Gorbar2}] where only the first $m=0$ electron state and $m=-1$ hole state were included. For increasing value of the charge, $<\hat{J}_z>$ decreases and deviates from the single impurity value. This is due to the fact that for increasing charge the angular symmetry gets more severely broken.

In Fig. 8 we show the region around the first anticrossing of Fig. 2(a) (between points (1) and (2)) and calculated the total angular momentum expectation value shown in color. Before the anticrossing the electron state has a positive angular momentum and the corresponding hole state has an equal angular momentum but with opposite sign and the value of $<\hat{J}_z>$ is clearly close to that of the $m=0$ (electron) and $m=-1$ (hole) state as discussed in the previous paragraph. When the states approach each other they start to hybridize into two new states and $\mid <\bar{J}_z>\mid$ decreases until it becomes zero at the point of anticrossing $Z\alpha\approx 0.94$. At anticrossing the probability density is equally distributed over the two charges (see middle bottom panel in Fig. 8) and the two states essentially behave as neutral, i.e. they are made up of half electron and half hole similar as a Majorana state. Next to that Fig. 8 again shows unambiguously how the energy levels continue regardless of the anticrossing and adds further support to the arguments made in the previous section. Around the anticrossing the wave functions change localization from one impurity to the other with increasing impurity charge as clearly shown in the bottem panel of Fig. 8. This effect was also mentioned in Refs. [\onlinecite{Gorbar}] and [\onlinecite{Gorbar2}]. 

\begin{figure}[h!!!]
\includegraphics[scale=0.43]{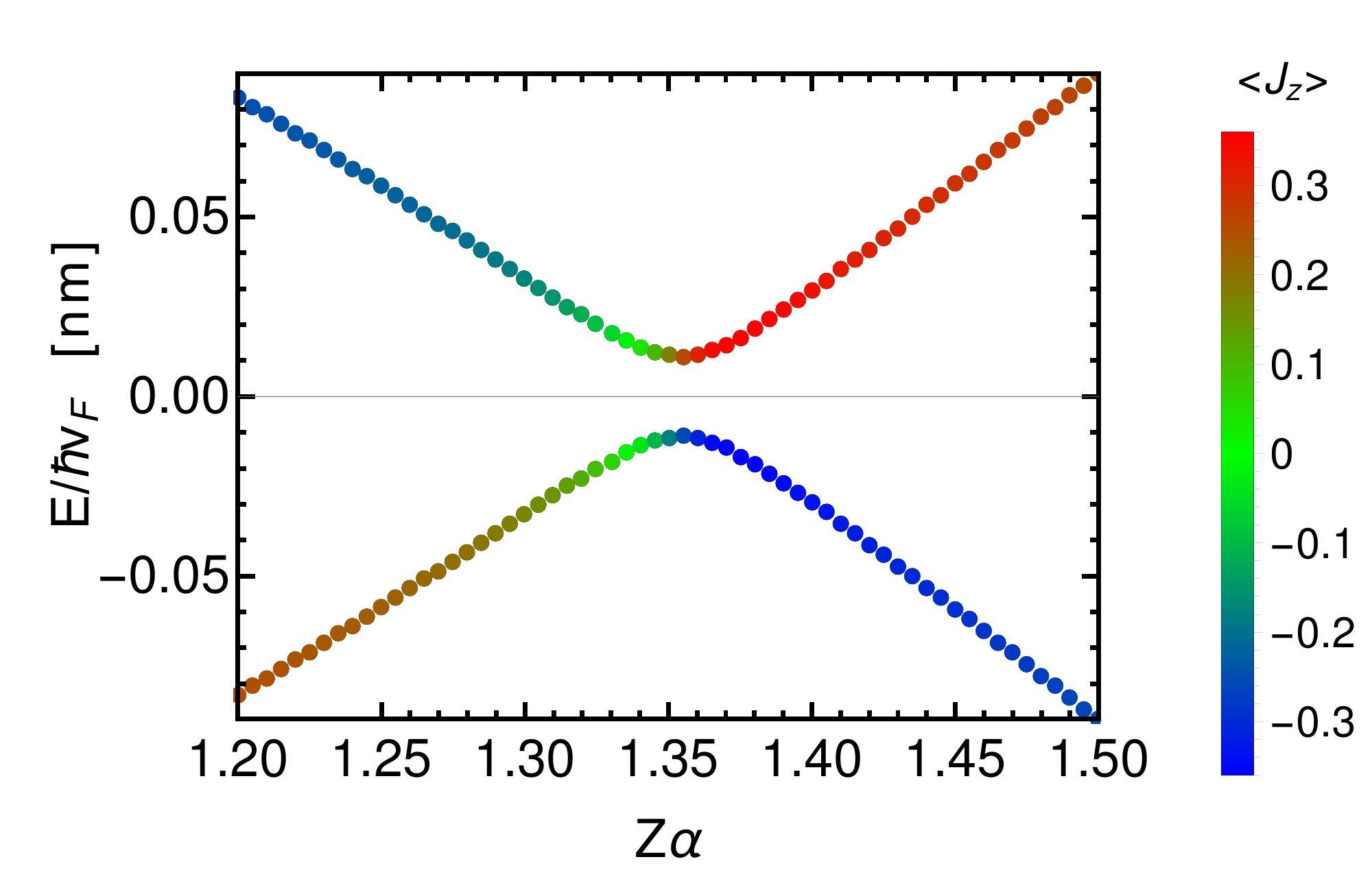}
\caption{Close up of the second anticrossing seen in Fig. 2(a). For each point we calculated the expectation value of the angular momentum operator $<\hat{J}_z>$.}
\end{figure}

In Fig. 9 we plot $<\hat{J}_z/\hbar>$ for the second anticrossing in Fig. 2(a) which occurs around $Z\alpha\approx 1.35$. While the energy dependence as function of the charge is very similar to that of the first anticrossing the evolution of $<\bar{J}_z>$ is very different. The electron state before the anticrossing has a negative expectation value compared to the positive one for the first anticrossing suggesting that it is related to the $m=-1$ level rather then the $m=0$ level as in the first anticrossing. This is also confirmed by the single impurity results shown in Fig. 3 where the second single impurity state is a $m=-1$ state. With increasing $Z\alpha$ we find that $\mid<\bar{J}_z>\mid$ decreases as in Fig. 8 but becomes zero before the point of closest approach. However, just as in Fig. 8 the continuation of the energy level, regardless of the anticrossing, can be clearly observed. The fact that the total angular momentum operator becomes zero before and not at the point of anticrossing is a consequence of the breaking of angular symmetry which is larger for the second anticrossing. 

For the densities shown in Fig. 5 we find the following values for the total angular momentum operator expectation value: $\{(8),-0.19\}$, $\{(9),-0.65\}$, $\{(10),-0.71\}$. This value is fairly close to the expectation value $<J_z>/\hbar=-0.5$ of a single impurity $m=-1$ state. This confirms our argument made in the previous section that this state is related to a $2s$ single impurity state.  

\section{Atomic collapse states in the continuum}
However, in order to get conclusive information about the atomic collapse it is essential to investigate what happens when states enter the continuum. In this section we show that the diving of the series of anticrossings persists as a peak in the local density of states (LDOS) providing a clear signature of atomic collapse in a dipole field. 

In the introduction it was already mentioned that the story of atomic collapse contains two parts: 1) the creation of an electron-hole pair, and 2) redistribution of the bound state over the continuum states as a resonance, visible as a peak in the LDOS in the negative continuum close to the impurity. As Pomeranchuk noted pair creation will not always occur while the peak in the LDOS in the negative continuum will always be observable and is the true signature for atomic collapse [\onlinecite{Pomeranchuk}]. 

Also from an experimental point of view the LDOS is interesting to calculate since it can be directly measured using an STM tip. Regarding atomic collapse it is most interesting to probe the LDOS at one of the impurity sites. In Fig. 10 we calculated the LDOS at the positively charged impurity for a dipole consisting of two charges separated by 10 nm from each other. We considered a finite square graphene flake of area 4900 $\text{nm}^2$ which corresponds to about 250.000 carbon atoms. A Gaussian broadening of $2.6 \text{ meV}$ was used for the energy levels. The edges of the positive and negative continuum are shown by the dashed lines. Comparing the LDOS with the spectrum in Fig. 2(a) we see clear differences, in Fig. 10 the LDOS of the diving states is larger than those of the rising states. This is due to the fact that we probe the positively charged impurity showing only the electron states. If we would probe the negatively charged impurity we would observe a similar spectrum but reflected around the $E=0$ axis.

From Fig. 10 we unambiguously observe how the energy levels dive into the continuum as function of $Z\alpha$ regardless of the anticrossings. This again confirms our previously made arguments. From Fig. 10 we also observe how the series of anticrossings enters and persists in the continuum where they form a clear and distinct peak in the LDOS. Just as in MLG this peak represents the redistribution of the bound state over the continuum states [\onlinecite{Dean}, \onlinecite{Vasilopoulos}]. The peaks visible in the LDOS inside the continuum will be a clear signature for the atomic collapse in a dipole field.

In Fig. 11 we plot the LDOS but now for charges located at a distance of $d=3$ nm from each other. Compared with Fig. 10 we see that both the descending and rising states have comparable LDOS at the positively charged impurity. This is due to the fact that the charges are located close to each other causing part of the probability density to leak to the negatively charged impurity (see Fig. 7). 

\begin{figure}[h!]
\includegraphics[scale=0.68]{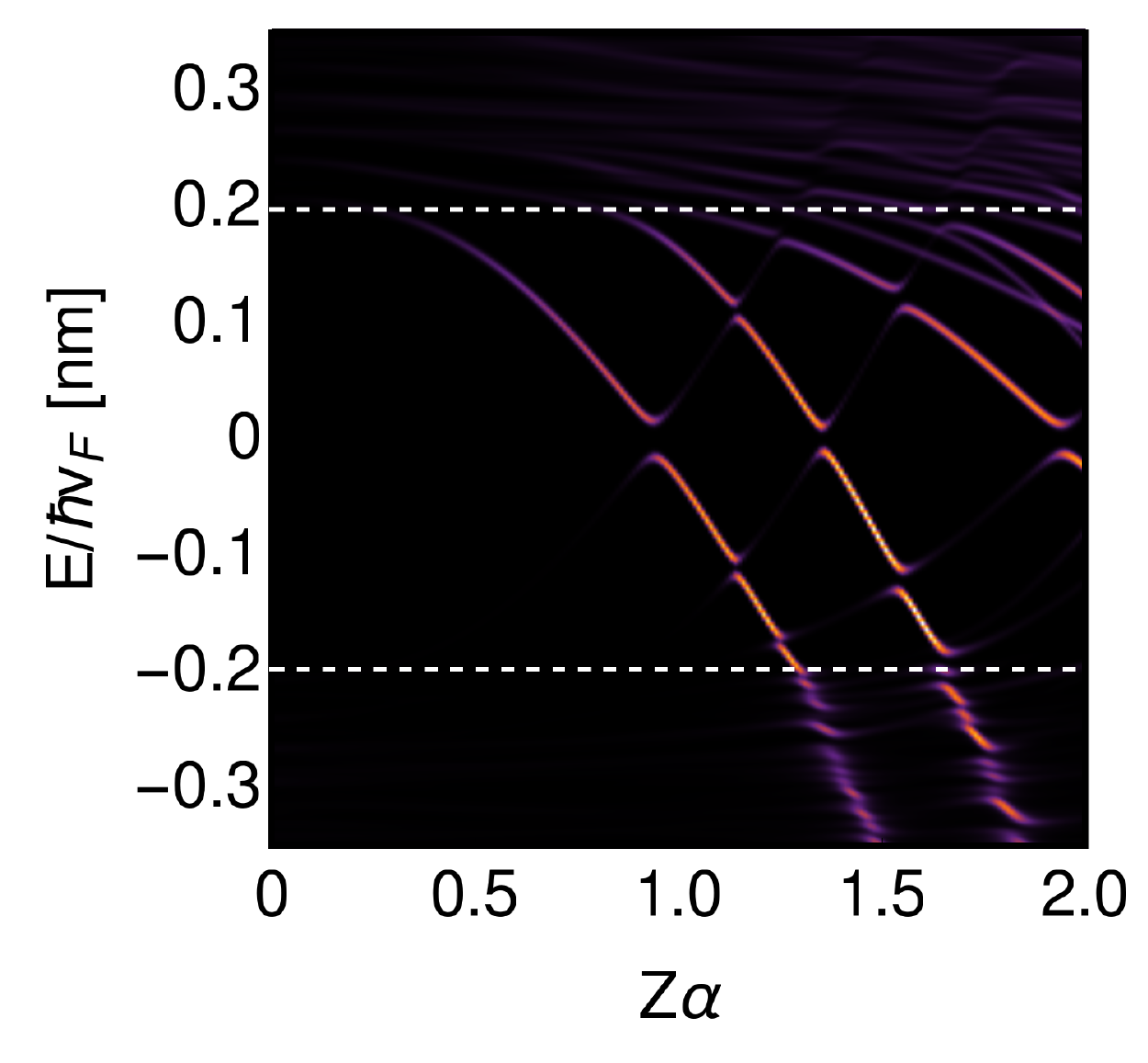}
\centering
\caption{(Color online) Local density of states calculated at the positively charged impurity as function of the impurity charges for $d=10$ nm. Black represents zero local density of states while yellow represents a high density of states. The dashed horizontal lines denote the gap edges.}
\end{figure}

\begin{figure}[h!]
\includegraphics[scale=0.68]{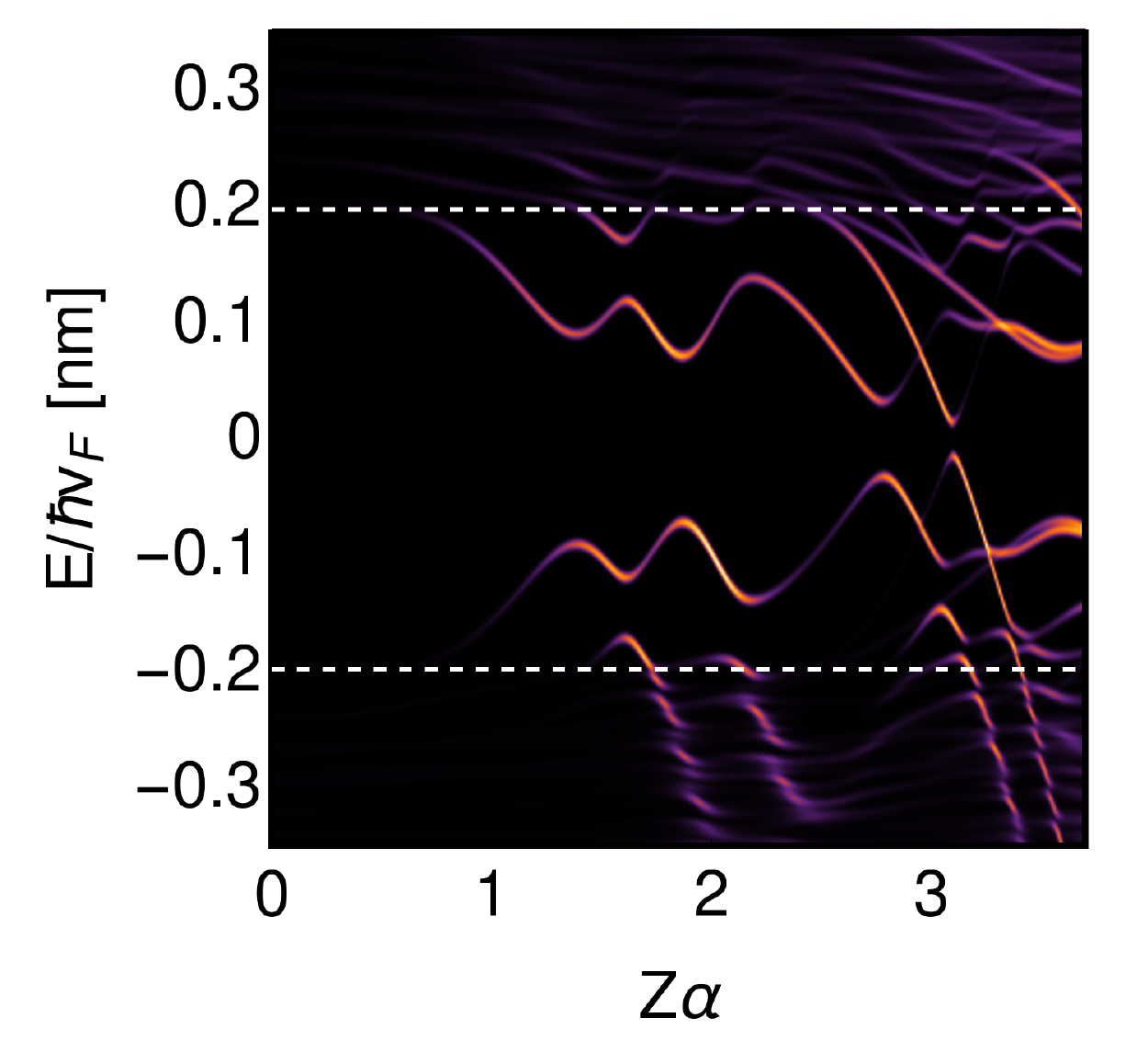}
\centering
\caption{(Color online) )The same as Fig. 10 but now for $d=3$ nm. }
\end{figure}

In Fig. 11 the state with sharp anticrossings that appears for $Z\alpha>2.5$, and also shown in Fig. 6, can be clearly observed. This state enters the continuum around the charge $Z\alpha=3.3$ and clearly persists after entering the continuum as a series of sharp anticrossings (note that in the limit $R\rightarrow\infty$ the series of anticrossings in the continuum will become located closer to each other to form a peak in the LDOS inside the continuum). This sharper series of anticrossings should be a clear signature to look for in experiments. 

It is also interesting to study the charge for which the first state enters the continuum as function of the distance between the impurities. This is analogous to the critical charge for atomic collapse for a single impurity (point for which the first state dives into the continuum). This critical charge is shown in Fig. 12 by the red symbols. The functional behavior can be well approximated (red curve) by the function $Z_c\alpha=1.1+1.76d^{-0.98}$ ($d$ is expressed in nm), showing almost perfect $1/d$ behavior. It is clear that the critical charge decreases with increasing distance. This can be understood from the fact that when the impurities come closer the strength of the Coulomb potential decreases softening the dependence on the charge. For large inter-impurity distances the critical charge converges to the critical charge of the single impurity case, $Z\alpha=1.1$ (see Fig. 3). When the inter-impurity distance becomes comparable to the Compton wavelength $R_\Delta=\hbar v_F/\Delta=5\text{ nm}$ the critical charge starts to depend more strongly on the inter-impurity distance. This can be understood from the fact that when the Compton wavelength is comparable to the inter-impurity distance the overlap between the charges increases enhancing the interaction between the electron and hole states. In Fig. 12 we show also the charge for which the first anticrossing occurs, that is the point where the electron-hole pair is created as suggested in Refs. [\onlinecite{Gorbar}] and [\onlinecite{Gorbar2}]. Again the functional behavior was close to $1/d$ giving the fit $Z_c\alpha=0.78+1.95d^{-1.08}$.  Note that the exact values of the critical charge will be sensitive to the value of the regularization parameter, see appendix A.

\begin{figure}[t]
\includegraphics[scale=0.45]{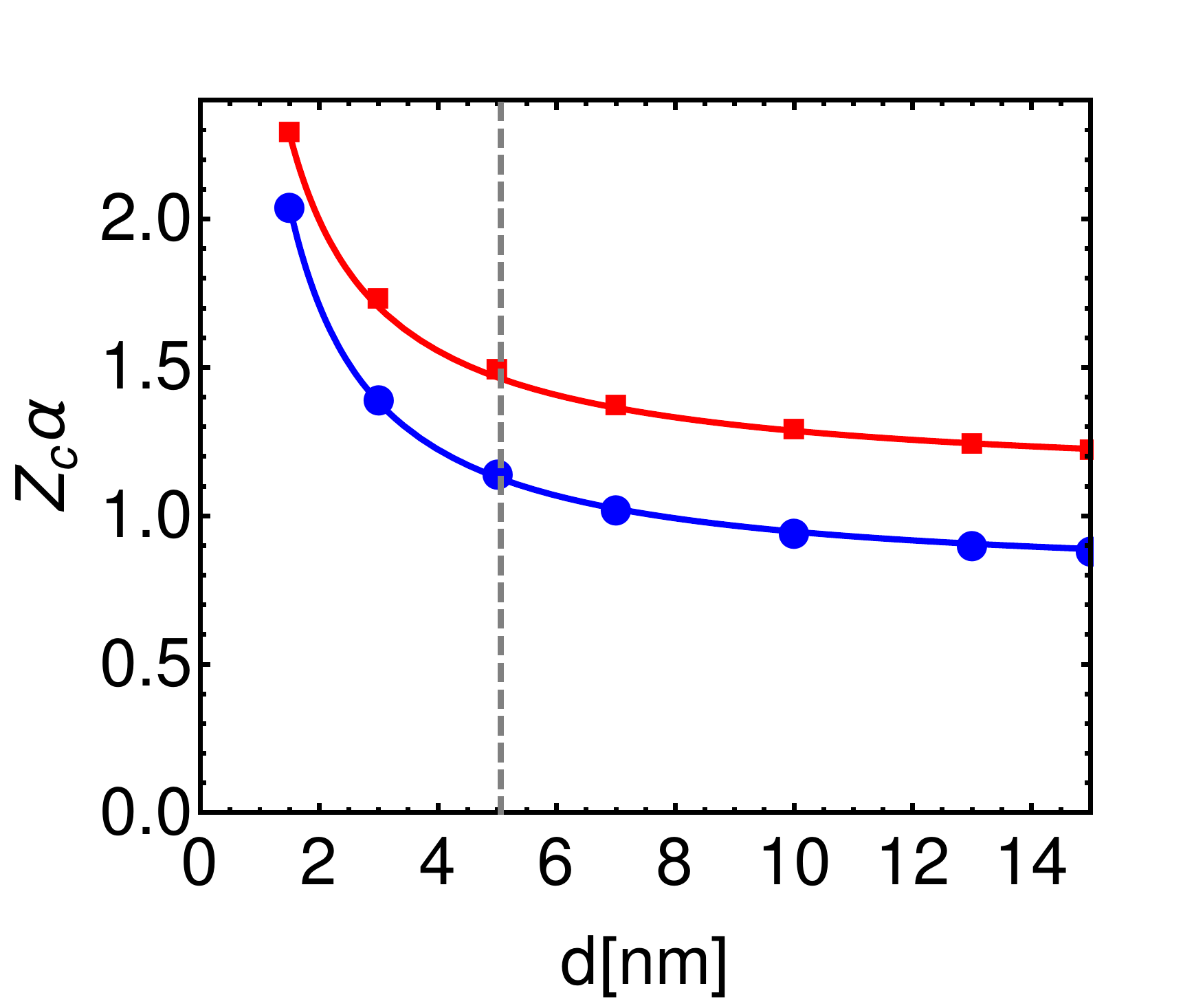}
\centering
\caption{(Color online) The charge at which the first anticrossing occurs (blue) and the charge for which the series of anticrossings enters the continuum (red) as function of the distance between the two charges. The red and blue curves fits with the function $Z_c\alpha=c+ax^b$, parameters are given in the text. The vertical dashed curve is where $d=R_\Delta$ the Compton wavelength.}
\end{figure}

The correspondence of our results with those of a recent publication [\onlinecite{Abdullah}] in which the energy levels of a quantum dot created by a blister was studied is remarkable. Similar anticrossings between electron and hole states were predicted in that case. However no attractive/repulse electrostatic potential, as in the dipole system, was present which is essential in the atomic collapse story. This is reflected in the fact that the anticrossings they observed never enter into the continuum, they approach the continuum but never reach it. The only common factor of the dipole and blister system is the fact that electron (conduction band) and hole (valence band) states switch positions so that an electron state is filled and a hole state emptied. Thus the fact that the levels hybridize and hence anticross \textit{inside the gap} is a direct consequence of the symmetry overlap between the states and is not related to atomic collapse. In accordance with atomic collapse in the single impurity system we find that the signature of atomic collapse in the dipole system is the appearance of the anticrossings in \textit{the negative/positive continuum} seen as a peak in the LDOS.

\section{Conclusions}
In this paper we investigated the full spectrum of a physical dipole located on the surface of a gapped graphene system within the continuum approximation. 

We investigated the evolution of the spectrum as function of the charge of the dipole and the inter-charge distance. By studying the probability density we showed that electron states located mostly at the positively charged impurities dive towards the negative continuum as a distinct series of anticrossings. Similarly hole states are mostly located at the negatively charged impurity dive as a series of anticrossings towards the positive continuum. From this result we can conclude that the diving of the states into the continuum is gradually replaced by a series of anticrossings that dive into the continuum in a dipole field. It is the appearance of this series of anticrossings \textit{inside the continuum}, seen as a peak in the LDOS, that is the signature of atomic collapse in a dipole field. with increasing flake size the anticrossings in the continuum will be located closer to each other.

When the distance between the impurities is increased the anticrossings become less and less pronounced showing a gradual evolution from the dipole to the single impurity case. We supported this observation by calculating the LDOS at the position of one of the impurities. The result of this calculation showed that the diving of the energy levels persist into the continuum and can be observed as a peak in the LDOS. The peaks visible after a state enters the continuum is the signature of atomic collapse in a dipole field. The calculated LDOS, and the atomic collapse states in the continuum,should be directly measurable in experiments. 

We also calculated the dependence of the critical charge for atomic collapse (the charge for which the first series of anticrossings enters the continuum analogous to MLG) as function of the distance between the impurities. The critical charge increases with decreasing inter-charge distance. 

Next to the atomic collapse phenomenon in the dipole system we also investigated general features of the energy spectrum. We investigated for example the influence of the regularization parameter on the spectrum and showed that the results only change quantitatively. 

We also noted a state diving towards the continuum exhibiting sharper anticrossings (i.e. with smaller anti-crossing gaps) and which was much more localized at one of the impurities compared to the other states. The probability density of this state is very asymmetric in the direction of the dipole axis which should be a clear signature in the experimentally measured LDOS. This state has not been observed and characterized in previous studies.

We also calculated the expectation value of the total angular momentum operator and investigated how the total probability density of the state changes around the anticrossings. We found that at the first point of anticrossing the two energy levels behave as neutral states, i.e. equally distributed over the two charges with zero total angular momentum.  

From our results we can now understand why in the literature different conclusions are reached with regard to the atomic collapse in a dipole field. In Refs. [\onlinecite{Gorbar}] and [\onlinecite{Gorbar2}] only the lowest electron and highest hole state were studied making it impossible to get the full picture of the states diving towards the continuum. On the other hand in Refs. [\onlinecite{Martino}] and [\onlinecite{Martino2}] the dipole approximation was employed. Because of this assumption they were not able to study the evolution of the spectrum as function of the inter-charge distance which is essential to understand the evolution of atomic collapse from a single impurity to the one in a dipole fields. Next to that extraction of the wave functions is very difficult with the method used in Ref. [\onlinecite{Martino}] making it even more difficult to understand the physics governing this system. 

In the single impurity system very good qualitative agreement between the experimental and theoretical predictions were observed [\onlinecite{crommie}, \onlinecite{andrei0}]. Keeping this in mind we can safely assume that our results should be at least qualitatively clearly observable in an experimental situation.

In conclusion, in this study we calculated for the first
time the full spectrum generated by a
dipole on gapped graphene without the use of any approximations. Our approach enabled us to study the spectrum of a dipole in a more profound and detailed way as compared to previous publications which allowed us to understand why different conclusions were reached in the literature. We showed that their is a continuous evolution from atomic collapse for single impurities to atomic collapse in a dipole field: states diving and hybridizing with the continuum are gradually replaced by a series of anticrossings representing a continuation of the single impurity behavior. Our results are complemented by the calculation of the experimental observable LDOS providing clear signatures to look for in experiments.

\acknowledgements

We thank Matthias Van der Donck for fruitful discussions. This work was supported by the Research Foundation of Flanders (FWO-V1) through an aspirant research grant for RVP and a postdoctoral grant for BVD. 
\newline

\appendix

\section{Influence of regularization}

\begin{figure}[h!!]
\includegraphics[scale=0.4]{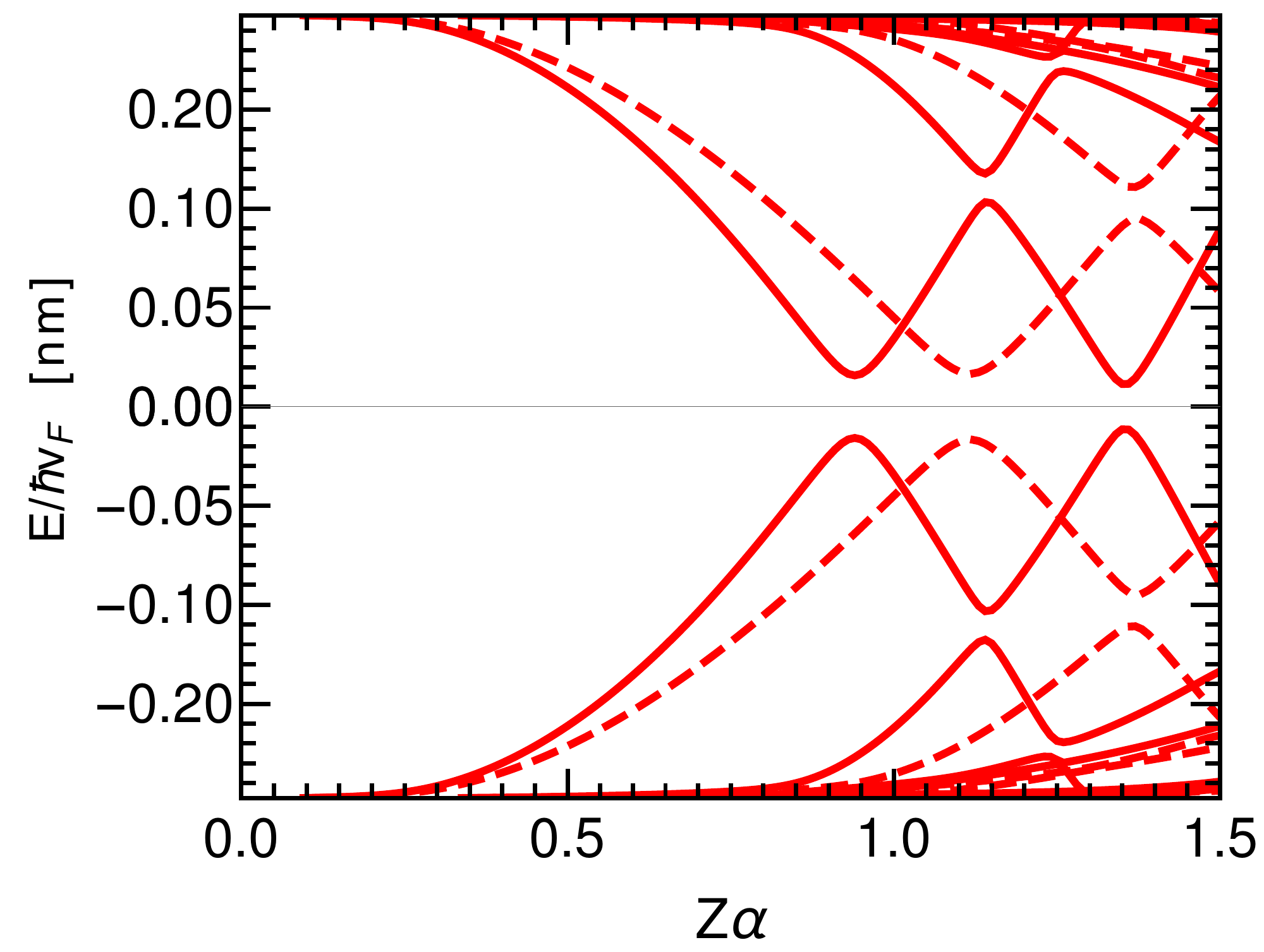}
\centering
\caption{Spectrum as function of the charge strength for a seperation distance $d=10$ nm for a dipole at $r_0=0.4$ nm (solid curves) and $r_0=0.7$ (dashed curves) from the graphene sheet.}
\end{figure}

As mentioned in the introduction regularisation is necessary to solve the problem for all values of the charge of the impurities [\onlinecite{Zarenia}]. This is because a point size impurity is singular leading to the so called fall-to-center phenomenon. In the latter case the problem can only be solved for charges $Z\alpha<0.5$, and to solve the problem beyond this charge proper regularization is required. In general this regularization can be done in different ways. One way is to take into account the finite size of the impurity analogous to the finite nucleus size in relativistic physics [\onlinecite{Dean}, \onlinecite{Vasilopoulos}]. Another way is to consider the charge at a certain distance from the graphene sheet [\onlinecite{andrei}]. In this paper we opted for the second kind of regularization which is given by formula (2) and is closely related to the experimental set up of Ref. [\onlinecite{crommie}].

In Fig. 13 the spectrum is shown for two different regularization parameters $r_0=0.4$ nm (solid curves) and $r_0=0.7$ nm (dashed curves). Notice, that the change in the size of the regularization parameter changes the spectrum quantitatively but not qualitatively. Increasing the regularization distance the charge for which the first anticrossing occurs shifts to higher values. This can be explained from the fact that increasing the regularization parameter weakens the strength of the Coulomb potential. 
\newline

\end{document}